\documentclass[apjl]{emulateapj}
\usepackage{apjfonts}

\newcommand{\cmjj}{\mbox{${\rm cm^{-2}}$}}
\newcommand{\hI}{\mbox{${\rm H\,I}$}}

\newcommand{\lya}{\mbox{${\rm Ly}\alpha$}}

\newcommand{\ha}{\mbox{${\rm H}\alpha$}}
\newcommand{\apg}{\gtrsim}
\newcommand{\apll}{\lesssim}
\newcommand{\etal}{\ensuremath{\mbox{et~al.}}}
\newcommand{\hmsol}{\mbox{$h^{-1}\,{\rm M}_\odot$}}
\newcommand{\ibid}{\underline{\makebox[0.5in]{}}.}
\providecommand{\kms}{\,\ensuremath{\rm{km\,s}^{-1}}}

\shorttitle{Baryon Content of Dark Matter Halos}
\shortauthors{Chen \& Tinker}

\begin{document}

\slugcomment{Accepted for Publication in the Astrophysical Journal}

\title{The Baryon Content of Dark Matter Halos: Empirical Constraints from Mg\,II Absorbers}

\author{Hsiao-Wen Chen and Jeremy L.\ Tinker} 
\affil{Dept.\ of Astronomy \& Astrophysics and 
Kavli Institute for Cosmological Physics, 
University of Chicago, Chicago, IL, 60637, U.S.A. \\ 
{\tt hchen@oddjob.uchicago.edu}}

\begin{abstract}

   We study the extent and covering fraction of cool baryons around
galaxies of different luminosity and mass, based on a survey of Mg\,II
$\lambda\lambda\,2796, 2803$ absorption features near known galaxies.
The initial sample consists of 13 galaxy and absorber pairs and 10
galaxies that do not produce Mg\,II absorption lines to within
sensitive upper limits.  The redshifts of the galaxy and absorber
pairs range from $z = 0.2067$ to 0.892 with a median of $z = 0.3818$.
We find that galaxies at larger impact parameters produce on average
weaker Mg\,II absorbers.  This anti-correlation is substantially
improved when accounting for the intrinsic luminosities of individual
galaxies.  In addition, there exists a distinct boundary at
$\rho=R_{\rm gas}$, beyond which no Mg\,II absorbers are found.  A
maximum likelihood analysis shows that the observations are best
described by an isothermal density profile and a scaling relation
$R_{\rm gas}=91\times (L_B/L_{B_*})^{(0.35\pm 0.05)}\ h^{-1}$ kpc (or
$69\ h^{-1}$ kpc at $W(2796)=0.3$ \AA) with a mean covering factor of
$\langle\kappa\rangle=80-86$ \%.  Together with the scaling relation
between halo mass and galaxy luminosity inferred from halo occupation
studies, this scaling of $R_{\rm gas}$ indicates that gas radius is a
fixed fraction of the dark matter halo radius.  We compare our results
with previous studies and discuss the implications of our analysis for
constraining the baryon content of galactic halos and for
discriminating between competing scenarios for understanding the
nature of the extended gas.

\end{abstract}

\keywords{cosmology: observations---intergalactic medium---quasars: absorption lines}

\section{INTRODUCTION}

  The Mg\,II $\lambda\lambda\,2796, 2803$ doublets are among the
absorption features commonly seen in the spectra of distant quasars
that are produced by intervening gaseous clouds along the quasar lines
of sight.  Their rest-frame absorption equivalent width $W(2796)$,
ranging from $W(2796)\apll 0.3$ \AA\ to $W(2796)>2$ \AA\ (e.g.\ Steidel
\& Sargent 1990; Nestor \etal\ 2005; Prochter \etal\ 2006a), is found
to represent the underlying gas kinematics (e.g.\ Petitjean \&
Bergeron 1990; Churchill \etal\ 2000).  Based on comparisons of the
abundance ratios of various associated ions, these Mg\,II absorption
transitions are understood to arise primarily in photo-ionized gas of
temperature $T\sim 10^4$ K (Bergeron \& Stas\'inska 1986; Hamann 1997)
and neutral hydrogen column density $N(\hI)=10^{18}-10^{22}$ \cmjj\
(Churchill \etal\ 2000; Rao \etal\ 2006).  In addition, surveys for
galaxies near the observed Mg\,II absorbers have often uncovered
luminous galaxies at projected distances ($\rho \apll 50\ h^{-1}$ kpc)
and velocity separations ($\Delta\,v \le 250$ \kms) from the absorbers
(Bergeron 1986; Lanzetta \& Bowen 1990, 1992; Steidel \etal\ 1994;
Zibetti \etal\ 2005; Nestor \etal\ 2007; Kacprzak \etal\ 2007).  The
relatively large associated $N(\hI)$ and the small separation between
Mg\,II absorbers and galaxies along common lines of sight indicate
that Mg\,II absorbers may offer a sensitive probe of photo-ionized halo
gas around galaxies at redshifts from $z=0.3$ to $z=2.3$ in the
optical spectral window.

  Understanding the physical origin of these Mg\,II absorbers bears
significantly on all efforts to apply their known statistical
properties for constraining the baryon content of dark matter halos on
different mass scales (Tinker \& Chen 2008).  Various theoretical
models have been developed in the past that describe the nature of the
absorbers in the context of gas accretion, including ram pressure
stripped gas from accreted satellites (Wang 1993), gravitationally
bound cold gas in halo substructures (e.g., Sternberg \etal\ 2002),
and condensed cold clouds in a thermally unstable hot halo (Mo \&
Miralda-Escude 1996; Maller \& Bullock 2004; Chelouche \etal\ 2007).

  Insights to the origin of Mg\,II absorbers can be obtained from
their clustering amplitude on large scales ($\apg\,1\,h^{-1}$ Mpc).
The clustering of the absorbers is a consequence of dark matter halos
in which they are found.  Measurements of the galaxy and absorber
cross--correlation function offer a means of quantifying the mean halo
mass of the absorbers. High-mass halos are expected to be highly
clustered, while low mass halos on average have weaker clustering
strength.  The cross--correlation function of Mg\,II absorbers and
luminous red galaxies (LRGs) have been measured at $\langle
z\rangle=0.6$ by Bouch\'e \etal\ (2006), who found that Mg\,II
absorbers of $W(2796)>0.3$ \AA\ cluster strongly with LRGs and that
weaker absorbers with $W(2796)=0.3-1.15$ \AA\ on average arise in dark
matter halos that are 10 times more massive than those of stronger
absorbers with $W(2796)=2-2.85$ \AA.

  Both the observed clustering amplitude and the inverse correlation
between the mean halo mass and absorber strength are difficult to
interpret.  For example, the frequency distribution function of Mg\,II
absorbers show that there are on average 10 times more $W(2796)=1$
\AA\ absorbers than those of $W(2796)=2$ \AA\ (e.g.\ Nestor \etal\
2005; Prochter \etal\ 2006a) which, when combined with the clustering
measurements, implies that the majority of absorbers reside in
massive, highly biased halos.  At the same time, we expect that
observed Mg$^+$ ions originate primarily in photo-ionized gas of
temperature $T\sim 10^4$ K and the halo gas in massive dark matter
halos becomes too hot for abundant Mg$^+$ to survive.

  In Tinker \& Chen (2008; hereafter TC08), we developed a new
technique that adopts the halo occupation framework for understanding
the origin of QSO absorption-line systems.  Specifically, the
technique adopts a model density profile for the Mg$^+$ ions in
individual dark matter halos.  The ``cold'' baryon content of
individual dark matter halos, as probed by the presence of Mg$^+$, is
then constrained through matching the space density and clustering
amplitude of dark matter halos with the observed frequency
distribution function of Mg\,II absorbers and their clustering
amplitude.  Our model allows the possibility that a predominant
fraction of the gas in massive halos is shock heated to the virial
temperature of the halo and becomes too hot to host abundant Mg\,II
absorbing clouds.  Within the hot halo, we further allow the
possibility that some dense, cold clouds may penetrate through as seen
in high resolution cosmological simulations (Kravtsov 2003).  The
result of this halo occupation analysis is the probability function
$P(W|M_h)$ that characterizes, for each dark matter halo of $M_h$, the
total probability of finding a Mg\,II absorber of equivalent width
$W$.

  The halo occupation analysis shows that observations, including both
the Mg\,II absorber frequency distribution function and their
clustering amplitude, demand a rapid transition in the halo gas
content at $M_h^{crit}\sim 10^{11.5}$ \hmsol.  Below $M_h^{crit}$,
halos contain predominantly cold gas and therefore contribute
significantly to the observed Mg\,II statistics. Beyond $M_h^{crit}$,
the cold gas fraction is substantially reduced and presumably the halo
gas becomes too hot to maintain a large contribution to Mg\,II
absorbers.  In order to reproduce the observed overall strong
clustering of the absorbers and the inverse correlation between
$W(2796)$ and halo mass $M_h$, roughly 5\% of the gas in halos up to
$10^{14}$ \hmsol\ is required to be cold.  It is understood under our
model that {\it the $W(2796)$ vs.\ $M_h$ inverse correlation arises
mainly as a result of an elevated clustering amplitude of $W(2796)\apll
1$ \AA\ absorbers, rather than a suppressed clustering strength of
$W(2796)\apg 2$ \AA\ absorbers.}  The clustering amplitude of weak
absorbers is elevated because of the presence of cold streams in
massive hot halos.  The amount of cold gas in clusters is small,
therefore these halos can only contribute to weak absorbers.

  The initial results of our halo occupation analysis demonstrate that
combining known statistics of dark matter halos with a simple model
for their gas content can already reproduce the statistical properties
known for Mg\,II absorbers.  It provides a simple prescription for
populating baryons in dark matter halos that can be compared directly
to results from numerical simulations and offer insights for
understanding the physics of gas accretion in halos of all mass
scales.  However, some uncertainties remain in the halo occupation
analysis.

  First, a key parameter that constrains the distribution of Mg$^+$
ions in individual dark matter halos is their incidence rate,
$\kappa(M_h)$, per halo.  It specifies the total probability of
detecting an absorber in a halo of mass $M_h$.  In the halo occupation
analysis of TC08, we find that dark matter halos of
$M_h=10^{11.5-12.5}$ \hmsol\ on average have 100\% incidence rate of
Mg$^+$ ions, $\langle\kappa(M_h)\rangle=1$, at $R_{\rm gas}\le R_{\rm
200}/3$ for Mg$^+$ ions and $\langle\kappa(M_h)\rangle=0$ at larger
radii.\footnote{The halo size $R_{\rm 200}$ corresponds to the radius,
within which the enclosed mean density is $200\times\bar{\rho}_m$ and
$\bar{\rho}_m$ is the background density.  It is motivated by
estimates of the virial radius both from the spherical collapse model
and from $N$-body simulations that predict $\approx
180\times\bar{\rho}_m$.}  In halos of lower masses,
$\langle\kappa(M_h)\rangle$ declines sharply.

  While a 100\% incidence rate shows that every dark matter halo hosts
a uniform gaseous halo of Mg$^+$ ions with size $R_{\rm gas}$, the low
$\langle\kappa(M_h)\rangle$ is more difficult to interpret because
$\langle\kappa(M_h)\rangle$ represents a mean value averaged over all
halos of mass $M_h$.  A low $\langle\kappa(M_h)\rangle$ may be a
result of only a small fraction of dark matter halos containing
extended distributions of Mg$^+$ ions or a result of a small covering
factor of Mg$^+$ in all halos.  In addition, we have assumed in our
initial analysis that the gaseous extent is related to halo mass
according to $R_{\rm gas}= R_{{\rm
gas}*}\,[M_h/(10^{12}\,\hmsol)]^{\beta}$, and $R_{{\rm gas}*}=50\
h^{-1}$ physical kpc and $\beta=1/3$.  This scaling relation follows
the theoretical expectation between virial radius and halo mass, but a
higher $R_{\rm gas}$ would naturally lower $\kappa$ for a fixed
frequency distribution function of Mg\,II absorbers.

  Second, the physical origin of the cold clouds probed by the Mg\,II
absorption transitions is ambiguous.  While the mass scale at which
halo gas is found to experience a transition from a cold-mode
dominated state to a hot-mode dominated state agrees well with the
expectation of theoretical models established to characterize the gas
accretion history in dark matter halos (e.g.\ Birnboim \& Dekel 2003;
Kere{\^s} \etal\ 2005; Dekel \& Birnboim 2006; Birnboim \etal\ 2007),
the observed cold-hot transition can also be interpreted as a
declining formation efficiency of cool clouds in a hot halo with
increasing halo mass (e.g.\ Mo \& Miralda-Escude 1996; Maller \&
Bullock 2004).  Both of these scenarios predict that halo gas is
primarily cold at $M_h\apll 10^{11.5}\,\hmsol$.  On the other hand,
Bouch\'e \etal\ (2006) have argued that the inverse correlation
between $W(2796)$ and clustering amplitude is suggestive of the
absorbers arising in non-virialized gas flows, such as starburst
winds.

  While dense clumps in starburst driven outflows are expected to
contribute to some fraction of the observed Mg\,II absorbers, an
important quantity to specify is the significance of this fraction.
If a large fraction of Mg\,II absorbers originate in cold dense clumps
in starburst winds, then one must also account for the more complex
star formation physics in efforts to apply known Mg\,II statistics for
constraining the gas content of dark matter halos.  
We defer to \S\ 5.3 for a more detailed discussion of the caveats in
interpreting the known properties of Mg\,II absorbers as evidence to
support their origin in starburst winds.

  Direct constraints on the covering fraction of Mg\,II absorbing gas
versus halo mass not only help to break the degeneracy between the
fraction of dark matter halos containing cold baryons and the extent
of cold gas in individual halos, in principle they also serve to
discriminate between the starburst wind scenario and gas accretion.
Under the starburst wind scenario, superwind driven outflows are
expected to proceed over a finite angular span along the minor axis
(e.g.\ Heckman \etal\ 1990; Veilleux \etal\ 2005), and result in only
partial covering of the halos.  In addition, because at a given epoch
only some fraction of galaxies are found at starburst or
post-starburst stages, we would expect a fraction of dark matter halos
to contain extended Mg\,II absorbing gas due to outflows.

  To obtain empirical constraints on the extent and covering fraction
of Mg$^+$ ions around galaxies of different luminosity and mass, we
have initiated a survey of Mg\,II absorbers around known galaxies at
small projected distances ($\rho\apll 100\ h^{-1}$ kpc) to a background
QSO.  Accounting the presence or absence of Mg\,II absorbers for an
unbiased sample of galaxies at different impact parameters yields a
statistical estimate of the gas covering fraction around field
galaxies.  A detailed comparison between the absorber strength and
galaxy properties further constrains the density profile and extent of
the gas.  The primary objectives of our study are (1) to improve the
uncertainties of our halo occupation analysis by obtaining empirical
measurements of $R_{\rm gas}$ and $\beta$, and (2) to examine whether
starburst driven outflows are the predominant mechanism for producing
the observed Mg\,II absorbers by obtaining an empirical constraint of
the gas covering factor $\kappa$.  Here we present the initial results
of our study.

  This paper is organized as follows.  In Section 2, we describe the
design of our experiment for constraining the extent and covering
fraction of Mg$^+$ ions.  In Section 3, we describe the imaging and
spectroscopic data available for our experiment.  In Section 4, we
examine the correlation between Mg\,II absorption strength and galaxy
properties, such as the impact parameter and luminosity.  We compare
our results with previous studies and discuss the implications of our
analysis in Section 5.  We adopt a $\Lambda$CDM cosmology,
$\Omega_{\rm M}=0.3$ and $\Omega_\Lambda = 0.7$, with a dimensionless
Hubble constant $h = H_0/(100 \ {\rm km} \ {\rm s}^{-1}\ {\rm
Mpc}^{-1})$ throughout the paper.

\section{EXPERIMENT DESIGN}

  Accurate estimates of the extent and covering fraction of Mg$^+$ ions
around galaxies require an unbiased survey of Mg\,II absorbers in the
vicinity of {\it known} galaxies.  Nearly all studies published in the
literature to date focus on surveys of galaxies that correspond to
known Mg\,II absorbers (e.g.\ Bergeron 1986; Lanzetta \& Bowen 1990,
1992; Steidel \etal\ 1994; Zibetti \etal\ 2007; Kacprzak \etal\ 2007),
with the exception of an on-going study described in Tripp \& Bowen
(2005).  Searching for galaxies that give rise to known Mg\,II
absorbers bears a large uncertainty due to contamination from
galaxies that are correlated with the true absorbing galaxies.  In
addition, these surveys based on known presence of Mg$^+$ ions do not
yield constraints on the incidence of these ions in galactic halos.

  We are conducting a survey of Mg\,II absorbers at the locations of
known galaxies close to the line of sight toward a background QSO.  We
identify galaxies at impact parameters $\rho\le r_0$.  The value of
$r_0$ is selected to allow identifications of galaxies at impact
parameters beyond the size of extended Mg\,II halos, in order to
constrain the extent and covering fraction of the gas.  At the same
time, we require $r_0$ to be small enough in order to minimize the
contamination of galaxies that do not give rise to the absorbers but
are merely correlated with the absorbing galaxies through clustering.

  To estimate the contamination rate of correlated galaxies as a
function of projected distance, we adopt the distribution of dark
matter halos identified in cosmological simulations as a guide.  We
use the $z=0.5$ output of a $400\ h^{-1}$ co-moving Mpc simulation
described in Tinker \etal\ (2007).  First, we identify for each random
line of sight dark matter halos that have $M_h \ge 10^{11}$ \hmsol\
and are located at $\rho\le R_{\rm 200}/3$ from the sightline.  From
TC08, we consider these halos likely to host a Mg\,II absorber along
the line of sight.  This is also consistent with the results from the
halo occupation distribution models of galaxies found in DEEP2 and
Sloan Digital Sky Survey from Zheng, Coil, \& Zehavi (2007; hereafter
ZCZ07).  These authors found that the minimum mass of a dark matter
halo that on average hosts a galaxy of $\apg\,0.1\,L_*$ is
$\approx\,2\times\,10^{11}$ \hmsol.  Next, for each potential Mg\,II
absorber, we search for neighboring halos that are at $|\delta\,v|\le
250$ \kms\ and projected distance less than $\rho$ from the absorber
and may therefore contaminate the search of the true absorber.
Finally, we repeat this exercise over 10,000 random lines of sight and
calculate the frequency of finding contaminating halos as a function
of $\rho$.

\begin{figure}
\begin{center}
\includegraphics[scale=0.4]{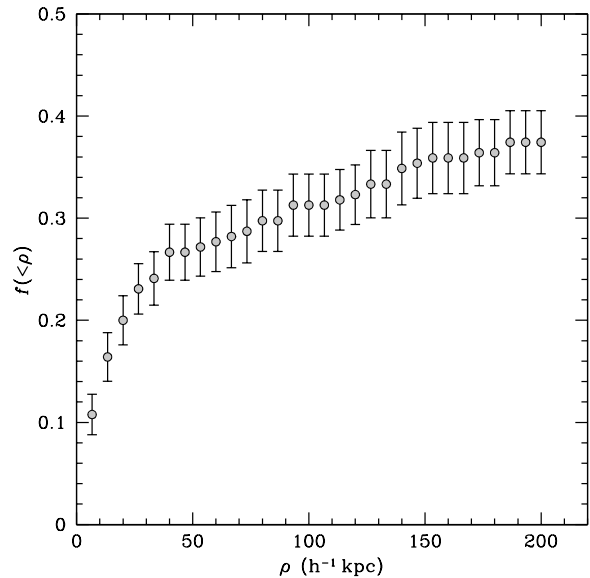}
\caption{The probability of finding more than one galaxy with
$|\delta\,v|\le 250$ \kms\ and at projected distances less than $\rho$
from an absorber. The errorbars represent the 68\% scatter that was
estimated using a jackknife resampling technique.}
\end{center} 
\end{figure}

  The result of our exercise is presented in Figure 1, where we show
the probability of contamination due to correlated galaxies versus
impact parameter.  The errorbars represent the 68\% scatter of the
mean from 16 jackknife subsamples.  Figure 1 shows that at $\rho>100\
h^{-1}$ physical kpc there is a $>$ 30\% chance that one would
mis-identify a correlated galaxy as the true absorber.  Combining the
contamination rate estimates with previous search reports that
indicate a mean gaseous extent of $\sim 50\ h^{-1}$ physical kpc for
luminous galaxies (Bergeron 1986; Lanzetta \& Bowen 1990, 1992;
Steidel \etal\ 1994), we carry out our study based on a sample of
galaxies that are within $r_0=100\ h^{-1}$ kpc from a background QSO.

\section{DATA}
 
  To study the extent and covering fraction of Mg$^+$ ions in galactic
halos, it requires survey data of both galaxies and absorbers along
common lines of sight toward background quasars.  We have selected
seven QSO fields for which galaxies data are already available, and
echelle spectra of the background QSOs have also been obtained either
from our own observations or through the ESO science data archive.  We
describe available QSO spectra and galaxy data separately in the
following sections.

\subsection{Echelle Spectra of QSOs}

  Echelle spectroscopic observations of the QSOs, PKS\,0122$-$0021
($z_{\rm QSO}=1.0765$), HE 0226$-$4110 ($z_{\rm QSO}=0.495$),
PKS\,0349$-$1438 ($z_{\rm QSO}=0.6163$), and PKS\,0454$-$2203 ($z_{\rm
QSO}= 0.5335$) were obtained using the MIKE echelle spectrograph
(Bernstein \etal\ 2003) on the Magellan Clay telescope in 2007
October.  The spectrograph contains a blue camera and a red camera,
allowing a full wavelength coverage from near-ultraviolet 3350 \AA\
through near-infrared 9400 \AA.  Depending on the brightness of the
QSO, the observations were carried out in a sequence of one to three
exposures of duration 600 s to 900 s each.  The mean seeing condition
over the period of integration was $0.7''$.  We used a $1''$ slit and
$2\times 2$ binning during readout, yielding a spectral resolution of
FWHM $\approx 13$ \kms\ at wavelength $\lambda=4500$ \AA\ and $\approx
16$ \kms\ at $\lambda=8000$ \AA.  The data were processed and reduced
using the MIKE data reduction software developed by Burles, Prochaska,
\& Bernstein\footnote{see
http://web.mit.edu/\char'176burles/www/MIKE/mike\_cookbook.html.}.
Wavelengths were calibrated to a ThAr frame obtained immediately after
each exposure and subsequently corrected to vacuum and heliocentric
wavelengths.  Flux calibration was performed using a sensitivity
function derived from observations of the flux standard EG131.  The
flux calibrated spectra have a mean $S/N>25$ per resolution element
across the entire spectral range.

  Echelle spectra of the QSOs, PKS1354$+$1933 ($z_{\rm QSO}=0.7200$),
PKS1424$-$1150 ($z_{\rm QSO}=0.8060$), and PKS1622$+$23 ($z_{\rm
QSO}=0.927$) were obtained using the Ultraviolet and Visual Echelle
Spectrograph (UVES; D'Odorico \etal\ 2000) on the VLT/UT2 telescope
and were retrieved from the ESO data archive.  Observations of
PKS1354$+$1933 (program ID 076.A$-$0860) and PKS1424$-$1150 (program
ID 075.A$-$0841) were obtained using a $1''$ slit and the dichroics
\#1 B346$+$R580 and \#2 B437$+$R860 in a sequence of two to three
exposures of duration 180s to 300 s each.  The data were binned
$2\times 2$ during readout.  Observations of PKS 1622$+$23 (program ID
069.A$-$0371) were obtained using a $1''$ slit and the dichroic \#1
B346$+$R580 in a sequence of three exposures of duration 4900 s each.
The data were also binned $2\times 2$ during readout.  The UVES
spectra were processed and calibrated using the standard data
reduction pipeline.  The combined spectra of the three QSOs cover a
spectral range from 3300 \AA\ to 6600 \AA, with a pixel scale of 0.035
\AA\ per pixel and $S/N>10$ per pixel at $\lambda>3700$ \AA.  The high
$S/N$ and high spectral resolution MIKE and UVES spectra together
allow us to search for Mg\,II absorption features at redshifts $0.2 <
z < 2.3$ to a limiting absorption strength $W(2796)=0.01$ \AA\ along
the lines of sight toward the seven background QSOs.

\begin{center}
\begin{small}
\begin{deluxetable*}{p{1in}cccccrc}
\tablewidth{0pc}
\tablecaption{Summary of the Echelle Spectroscopic Observations of Distant QSOs} 
\tabletypesize{\small}
\tablehead{ \colhead{Field} & \colhead{RA(J2000)} & \colhead{Dec(J2000)} & \colhead{$z_{\rm QSO}$} & \colhead{$V$} & \colhead{Instrument} & \colhead{Exptime} & \colhead{Spectral Coverage}}
\startdata
PKS\,0122$-$0021 \dotfill & 01:25:28.8 & $-$00:05:56 & 1.0765 & 16.4 & MIKE & $1200$ & 3350--9400 \nl 
HE\,0226$-$4110  \dotfill & 02:28:15.2 & $-$40:57:16 & 0.4950 & 15.2 & MIKE & $600$ & 3350--9400 \nl
PKS\,0349$-$1438 \dotfill & 03:51:28.5 & $-$14:29:09 & 0.6163 & 16.2 & MIKE & $1800$ & 3350--9400 \nl
PKS\,0454$-$2203 \dotfill & 04:56:08.9 & $-$21:59:09 & 0.5335 & 16.1 & MIKE & $2700$ & 3350--9400 \nl
PKS\,1354$+$1933 \dotfill & 13:57:04.4 & $+$19:19:07 & 0.7200 & 16.0 & UVES & $600$ & 3300--6800 \nl
PKS\,1424$-$1150 \dotfill & 14:27:38.1 & $-$12:03:50 & 0.8060 & 16.5 & UVES & $720$ & 3300--6600 \nl 
3C336            \dotfill & 16:24:39.1 & $+$23:45:12 & 0.9274 & 17.5 & UVES & $14,700$ & 3300--6600 \nl
\enddata
\end{deluxetable*}
\end{small}
\end{center}

\subsection{The Galaxy Sample}

  For the purpose of the study described in \S\ 2, we need to identify
a representative sample of galaxies at $\rho\apll 100\ h^{-1}$ kpc from
the background QSOs. Galaxies in six of the seven QSO fields listed in
Table 1 have been observed and analyzed by Chen \etal\ (1998, 2001a).
Specifically, galaxies in the fields around PKS\,0122$-$0021,
PKS\,0349$-$1438, and PKS\,0454$-$2203, PKS1354$+$1933, and
PKS1424$-$1150 were observed as part of the survey program to study
the gaseous extent of galaxies based on the presence or absence of
corresponding \lya\ and C\,IV absorption features in the spectra of
the background QSOs (Lanzetta \etal\ 1995; Chen \etal\ 1998, 2001a;
Chen \etal\ 2001b).  The authors carried out a redshift survey of
galaxies brighter than $R=21.5$ mag at angular distances $\theta <
1.3'$.  The survey depth enabled findings of galaxies with
luminosities greater than $0.01-1.0\ L_*$ and impact parameters less
than $100-350\ h^{-1}$ kpc at the redshifts $0.1\apll z_{\rm gal}\apll
0.6$.  The galaxy sample is therefore ideally suited for our present
study.

  Galaxies in the field of PKS\,1622$+$2352 were first published in
Steidel \etal\ (1997) as part of a program to determine the redshifts
of galaxies found in deep Hubble Space Telescope (HST) images that are
brighter than $R\approx 24.5$.  Images of individual galaxies were
also analyzed by Chen \etal\ (1998).  Galaxies in the field of
HE\,0226$-$4110 were observed in an on-going program to uncover
galaxies brighter than $R=23$ mag at $z_{\rm gal}<0.5$ for
understanding the cross-correlation between galaxies and O\,VI
$\lambda\lambda\,1031, 1037$ absorbers (Chen \etal\ 2008 in
preparation).  The two galaxy samples satisfy the selection criterion
of an unbiased galaxy survey in QSO fields that does not rely on known
knowledge of the presence of an Mg\,II absorber.  We therefore include
the galaxy data in our sample.  Images of the galaxies are presented
in Figure 2.  The data were obtained using the Wide Field and
Planetary Camera 2 (WFPC2) and the F702W filter (see also Chen \etal\
2001a) on board HST, with the exception of the images of galaxies in
the field around HE\,0226$-$4110 that were obtained using IMACS and
the $R$-band filter on the Magellan Baade telescope.

\begin{figure*}
\begin{center}
\includegraphics[scale=0.6,angle=0]{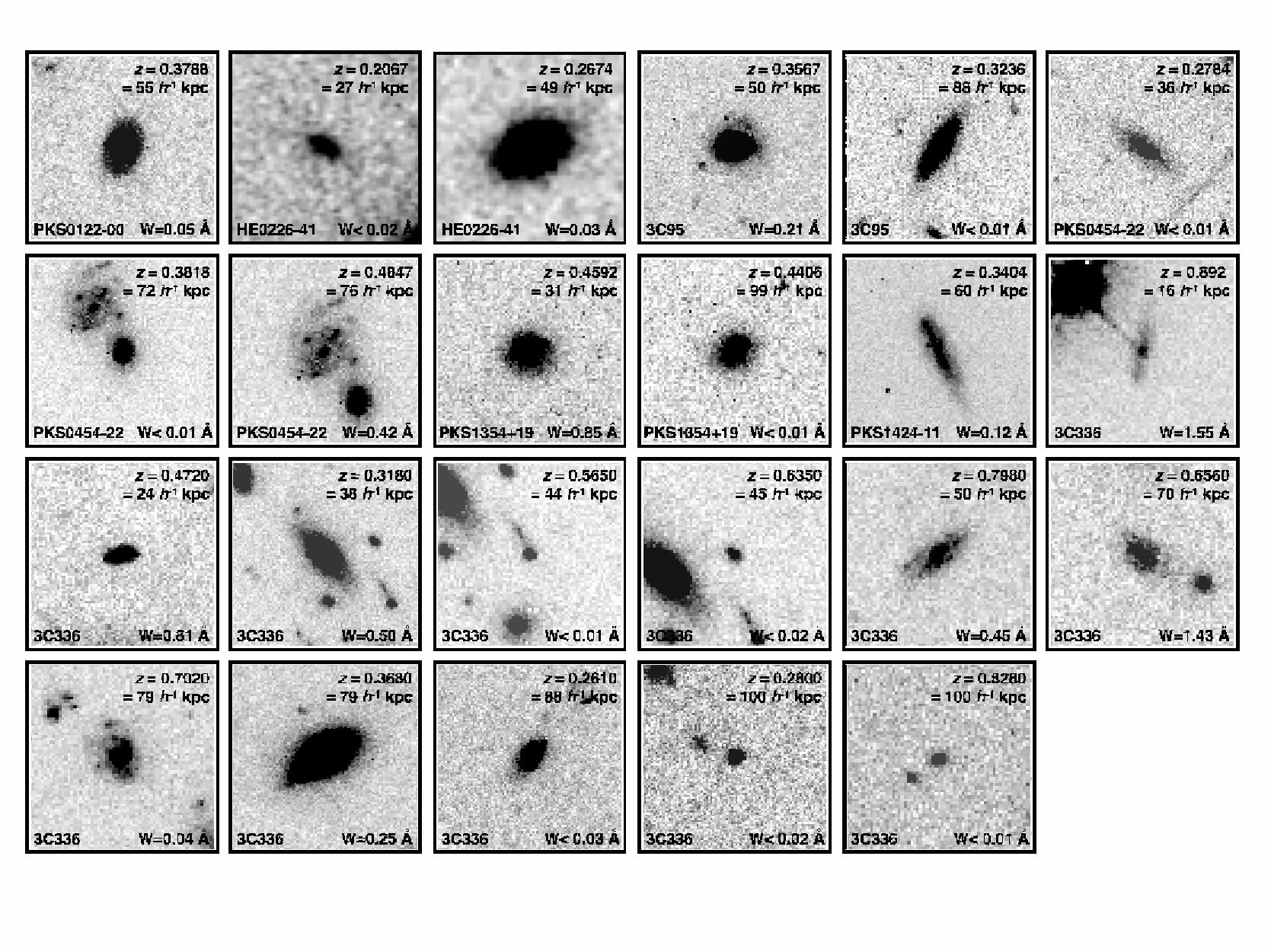}
\caption{Images of the 23 galaxies in our sample that are identified
at $\rho\apll 100\ h^{-1}$ kpc from the lines of sight toward a
background QSO.  With the exception of the two galaxies in the field
of HE\,0226$-$4110, the galaxy images were obtained using WFPC2 and
the F702W filter (see also Chen \etal\ 2001a). Images of the galaxies
in the field of HE\,0226$-$4110 were obtained using IMACS and the $R$
filter on the Magellan Baade telescope.  The spatial extent of each
image is roughly 25 $h^{-1}$ physical kpc on a side at the redshift of
the galaxy indicated in the upper right-corner.  The orientation of
each image is arbitrary.  The absorption strength of the corresponding
Mg\,II absorbers found in the spectra of the background QSO described
in \S\ 3.1 is indicated in the lower-right corner.  For
non-detections, we provide a 3-$\sigma$ upper limit to the rest-frame
absorption equivalent width.}
\end{center} 
\end{figure*}

  The sample of galaxies identified at $\rho\apll 100\ h^{-1}$ kpc from
the lines of sight toward seven background QSOs is presented in Table
2, where we list for each galaxy in columns (1)---(7) the field, Right
Ascension and Declination offsets from the QSO $\Delta \alpha$ and
$\Delta \delta$, redshift $z_{\rm gal}$, impact parameter $\rho$,
apparent $R$-band magnitude, and absolute $B$-band magnitude $M_B - 5
\log h$.\footnote{All magnitudes presented in this paper are in the
$AB$ system.}  Measurement uncertainties in $R$ and $M_B - 5 \log h$
were typically $0.2$ dex.  Our initial galaxy sample consists of 23
galaxies at redshifts from $z_{\rm gal} = 0.2067$ to $z_{\rm
gal}=0.892$ with a median of $z_{\rm gal} = 0.3818$\footnote{We note
that 12 of the 23 galaxies in our sample are from the field around
3C336.  These 12 galaxies have spectroscopic redshifts evenly
distributed from $z=0.261$ to $z=0.892$ with a median of $\langle
z\rangle=0.6$, covering a comoving distance of $\approx 1.4\ h^{-1}$
Gpc.  Therefore these galaxies are not part of a coherent structure.
There is no known galaxy clusters along the sightline toward 3C336 and
there is no physical mechanism that suggests that slight overdensities
on scales of 0.1-1 Gpc would affect gas properties within individual
halos.}.  We have excluded galaxies that are within $|\Delta\,v|<3000$
\kms\ from the background QSOs, where most of the gas is expected to
be highly ionized due to the QSO proximity effect (e.g.\ Bajtlik
\etal\ 1988).  The rest-frame $B$-band magnitudes of the galaxies
range from $M_B-5\,\log\,h=-15.9$ to $-20.3$ with a median of
$M_B-5\,\log\,h=-19.0$, corresponding to $0.5\,L_{B_*}$ for
$M_{B_*}-5\,\log\,h=-19.8$ that characterizes the blue galaxy
population at $z_{\rm gal}\sim 0.4$ (e.g.\ Faber \etal\ 2007).

\subsection{The Galaxy--Mg\,II Absorber Pair Sample}

  To establish a galaxy--Mg\,II absorber pair sample for the
subsequent analysis, we examine the echelle spectra of the background
QSOs and search for the corresponding Mg\,II absorption doublet at the
locations of the galaxies presented in \S\ 3.2.  Specifically, we
accept absorption lines according to a $3 \sigma$ detection threshold
criterion, which is appropriate because the measurements are performed
at a small number of known galaxy redshifts.  Next, we form galaxy and
absorber pairs by requiring a velocity difference\footnote{We note
that the galaxy at $z_{\rm gal}=0.4592$ in the field around
1354$+$1933 is $\approx 550$ \kms\ from the identified Mg\,II
absorber.  However, including the redshift measurement uncertainty of
300 \kms\ reported in Ellingson \etal\ (1991), we consider this pair a
match. } $\Delta\,v\le 250$ \kms.  Finally, we measure 3-$\sigma$
upper limits to the 2796-\AA\ absorption equivalent widths of galaxies
that are not paired with corresponding absorbers.  The procedure
identifies 13 galaxy and absorber pairs and 10 galaxies that do not
produce Mg\,II absorption lines to within sensitive upper limits.
Impact parameter separations of the galaxy and absorber pairs range
from $\rho = 16.3$ to $100.1 \ h^{-1}$ physical kpc with a median of
$\rho = 54.8 \ h^{-1}$ kpc.  The corresponding Mg\,II absorption
strength of each of the 23 galaxies in our sample is listed in columns
(8)---(9) of Table 2.

\begin{center}
\begin{deluxetable*}{lrrccccccc}
\tablewidth{0pc}
\tablecaption{Galaxies and Absorption Systems} 
\tabletypesize{\small}
\tablehead{ & \multicolumn{6}{c}{Galaxies} & & \multicolumn{2}{c}{Absorption Systems} \\
\cline{2-7}
\cline{9-10}
& \multicolumn{1}{c}{$\Delta \alpha$} & \multicolumn{1}{c}{$\Delta \delta$} & &
\multicolumn{1}{c}{$\rho$} & & \colhead{$M_B$} & & & \colhead{$W(2796)$} \\
\multicolumn{1}{c}{Field} & \multicolumn{1}{c}{(arcsec)} &
\multicolumn{1}{c}{(arcsec)} & \colhead{$z_{\rm gal}$} & \colhead{($h^{-1}$ kpc)} & \colhead{$R$\tablenotemark{a}} & 
\colhead{$-5 \log h$} & & \colhead{$z_{\rm abs}$} & \colhead{(\AA)} \\
\multicolumn{1}{c}{(1)} & \multicolumn{1}{c}{(2)} & \multicolumn{1}{c}{(3)} & 
\colhead{(4)} & \colhead{(5)} & \colhead{(6)} & \colhead{(7)} & & \colhead{(8)} & \colhead{(9)} }
\startdata
0122$-$0021 & $  -8.7$ & $ -12.3$ & $0.3788$ & $  54.8$ & $ 20.6$ & $-19.2$ & & $0.3791$ & $0.05\pm 0.01$ \\
0226$-$4110 & $  -7.9$ & $  -7.9$ & $0.2067$ & $  26.5$ & $ 22.2$ & $-16.4$ & & ... & $< 0.02$ \\ 
            & $  11.6$ & $ -12.5$ & $0.2674$ & $  49.0$ & $ 20.4$ & $-18.4$ & & $0.2678$ & $0.03\pm 0.01$ \\ 
0349$-$1438 & $  -9.4$ & $  10.8$ & $0.3567$ & $  50.1$ & $ 20.6$ & $-19.1$ & & $0.3572$ & $0.21\pm 0.01$ \\ 
            & $  11.7$ & $ -24.1$ & $0.3236$ & $  88.0$ & $ 20.0$ & $-19.1$ & & ... & $<0.01$ \\ 
0454$-$2203 & $  10.6$ & $   5.7$ & $0.2784$ & $  35.6$ & $ 21.1$ & $-18.0$ & & ... & $<0.01$ \\ 
            & $  -1.1$ & $ -18.0$ & $0.4847$ & $  75.8$ & $ 20.1$ & $-20.3$ & & $0.4834$ & $0.42\pm 0.01$ \\ 
            & $   0.3$ & $ -19.7$ & $0.3818$ & $  72.0$ & $ 20.8$ & $-19.1$ & & ... & $<0.01$ \\ 
1354$+$1933 & $   1.2$ & $   7.5$ & $0.4592$ & $  31.0$ & $ 21.0$ & $-19.3$ & & $0.4563$ & $0.85\pm 0.01$ \\ 
            & $ -21.6$ & $ -12.2$ & $0.4406$ & $  98.8$ & $ 20.9$ & $-19.1$ & & ... & $<0.01$ \\ 
1424$-$1150 & $  -0.2$ & $  17.6$ & $0.3404$ & $  59.8$ & $ 20.5$ & $-19.1$ & & $0.3415$ & $0.12\pm 0.04$ \\ 
3C336       & $   3.0$ & $  -0.0$ & $0.8920$ & $  16.3$ & $ 23.3$ & $-19.0$ & & $0.8909$ & $1.55\pm 0.01$ \\ 
            & $  -4.2$ & $  -3.8$ & $0.4720$ & $  23.6$ & $ 22.5$ & $-17.9$ & & $0.4716$ & $0.81\pm 0.01$ \\ 
            & $  -5.5$ & $   7.5$ & $0.6350$ & $  44.7$ & $ 24.1$ & $-17.1$ & & ... & $<0.02$ \\ 
            & $  -8.9$ & $   3.2$ & $0.7980$ & $  49.9$ & $ 22.4$ & $-19.5$ & & $0.7969$ & $0.45\pm 0.01$ \\ 
            & $  -3.3$ & $   9.1$ & $0.5650$ & $  44.1$ & $ 23.8$ & $-17.1$ & & ... & $<0.01$ \\ 
            & $  -6.7$ & $   9.7$ & $0.3180$ & $  38.3$ & $ 20.2$ & $-18.9$ & & $0.3172$ & $0.50\pm 0.01$ \\ 
            & $   1.6$ & $  14.2$ & $0.6560$ & $  69.6$ & $ 22.7$ & $-18.6$ & & $0.6556$ & $1.43\pm 0.01$ \\ 
            & $ -12.3$ & $   9.9$ & $0.7020$ & $  79.1$ & $ 21.9$ & $-19.6$ & & $0.7023$ & $0.04\pm 0.01$ \\ 
            & $   7.9$ & $ -17.0$ & $0.8280$ & $  99.7$ & $ 24.2$ & $-18.3$ & & ... & $<0.01$ \\ 
            & $ -21.3$ & $  -6.2$ & $0.3680$ & $  79.2$ & $ 19.9$ & $-19.9$ & & $0.3677$ & $0.25\pm 0.01$ \\ 
            & $ -24.1$ & $   5.3$ & $0.3680$ & $  81.0$ & $ 23.4$ & $-16.4$ & & ... & ... \\ 
            & $  25.5$ & $  17.6$ & $0.2610$ & $  87.5$ & $ 21.6$ & $-17.3$ & & ... & $< 0.03$ \\ 
            & $  12.7$ & $ -31.2$ & $0.2800$ & $ 100.1$ & $ 22.9$ & $-15.9$ & & ... & $< 0.02$ \\ 
\enddata

\tablenotetext{a}{The $R$ magnitudes presented here were measured in
HST/WFPC2 F702W filter for all galaxies but the two in the field
around HE0226$-$4110.  No filtered HST images are available for the
field of HE0226$-$4110.  The magnitudes were determined based on a
ground-based image obtained with the Magellan Baade telescope using
IMACS and the Johnson $R$ filter.}
\end{deluxetable*}
\end{center}

\section{THE EXTENT AND COVERING FACTOR OF Mg$^+$ IONS IN GALACTIC HALOS}

  In this section, we examine the correlation between galaxies and
Mg\,II absorbers identified along common lines of sight, using the
sample of 13 galaxy--Mg\,II absorber pairs and 10 galaxies at
$\rho\apll 100\ h^{-1}$ kpc that do not give rise to Mg\,II absorption
to a sensitive upper limit.  The choice of $r_0=100\ h^{-1}$ kpc in
selecting galaxies at $\rho\apll r_0$ for our analysis is adopted, in
order to minimize the contamination of galaxies that are correlated
with the true Mg\,II absorbing galaxies through clustering and to
maximize the fraction of physical galaxy and absorber pairs in our
sample.  In addition to the simulation results presented in \S\ 2,
this criterion is further supported by the stronger clustering signal
at small ($\rho<100\ h^{-1}$ kpc) impact parameters of LRGs (Bouch\'e
\etal\ 2004).  This strong signal is a natural consequence of
cross-correlating objects inside the same dark matter halos of the
LRGs.

  To examine the existence of a fiducial relationship between the
strength of Mg\,II absorption and galaxy impact parameter, we present
in Figure 3 the distribution of $W(2796)$ with respect to $\rho$.
Circles represent elliptical or S0 galaxies, triangles represent
early-type spiral galaxies, and squares represent late-type spiral
galaxies.  Galaxy morphological types are adopted from the
classifications of Chen \etal\ (1998, 2001a) that were based on a 2D
surface profile analysis of the high resolution HST/WFPC2 images.  We
have also classified the galaxy at ($-7.9'',-7.9''$) and the galaxy at
($11.6'',-12.5''$) from the QSO HE\,0226$-$4110 as a late-type and an
early-type spiral galaxy based on their respective spectral features.
Points with arrows in Figure 3 indicate $3 \sigma$ upper limits.

\begin{figure}
\begin{center}
\includegraphics[scale=0.45]{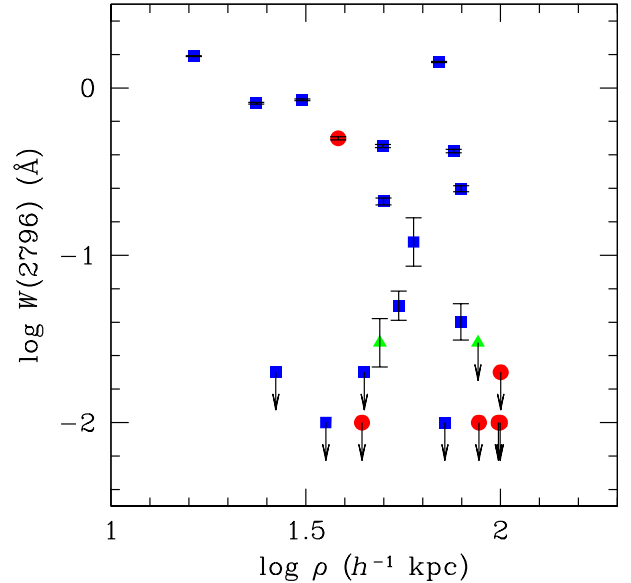}
\caption{Comparison of the corresponding rest-frame absorption
equivalent width $W(2796)$ versus galaxy impact parameter $\rho$ for a
sample of 23 galaxies.  Circles represent elliptical or S0 galaxies,
triangles represent early-type spiral galaxies, and squares represent
late-type spiral galaxies.  Points with arrows indicate $3 \sigma$
upper limits. }
\end{center} 
\end{figure}

  Figure 3 shows that there is a moderate anti-correlation between
$W(2796)$ and $\rho$, i.e.\ weaker Mg\,II absorption strength is
observed for galaxies at larger impact parameter.  We note that the
obvious outlier of a Mg\,II absorber with $W(2796)=1.43\pm 0.01$ \AA\
at $z=0.656$, currently assigned to a galaxy at $\rho=69.6\ h^{-1}$
kpc, is a known damped \lya\ absorber of $\log\,N(\hI)=20.36\pm 0.08$
(Steidel \etal\ 1997; Rao \etal\ 2006).  Although the galaxy satisfies
our sample selection criterion, it is unlikely that the galaxy at
$\rho=69.6\ h^{-1}$ kpc is physically related to the absorber due to
the large neutral gas content for two reasons.  First, all radio
detected sources in non-cluster environment from deep 21cm surveys
have coincident optical counterparts (e.g.\ Doyle \etal\ 2005).
Second, deep 21cm observations of the M\,31 halo reveal no H\,I clouds
of $N(\hI)>10^{20}$ \cmjj\ (e.g.\ Thilker \etal\ 2004).  This is also
roughly consistent with the expected contamination rate at this
projected separation presented in Figure 1.

  To assess the significance of this anti-correlation, we perform a
generalized Kendall test that accounts for the presence of
non-detections and find that the null hypothesis, in which $W(2796)$
is randomly distributed with respect to $\rho$, can be ruled out at a
$>97$\% confidence level.  However, the scatter of the $W(2796)$ vs.\
$\rho$ anti-correlation is clearly large.  Specifically, 10 of the 23
galaxies at $\rho\apll 100\ h^{-1}$ kpc do not produce a corresponding
Mg\,II absorber to a sensitive upper limit, implying a mean covering
factor of the Mg$^+$ ions $\langle\kappa\rangle\approx 57$ \% at
$\rho<100\ h^{-1}$ kpc.  At $\rho < 45\ h^{-1}$ kpc, we find
$\langle\kappa\rangle\approx 50$ \%, similar to the preliminary result
reported in Tripp \& Bowen (2005).

  We note, however, that the rest-frame $B$-band magnitudes of the
galaxies range from $M_B-5\,\log\,h=-15.9$ to $-20.3$ with a median of
$M_B-5\,\log\,h=-19.0$, corresponding to a range in intrinsic $B$-band
luminosity from $0.03\,L_{B_*}$ to $1.6\,L_{B_*}$ with a median of
$0.5\,L_{B_*}$ for $M_{B_*}-5\,\log\,h=-19.8$ at $z_{\rm gal}\sim 0.4$
(e.g.\ Faber \etal\ 2007).  Because more massive galaxies are expected
to reside in more massive halos and therefore possess larger gaseous
halos, we examine whether accounting for galaxy intrinsic luminosity
can improve upon the anti-correlation displayed in Figure 3.

  Following TC08, we adopt an isothermal profile of finite extent
$R_{\rm gas}$ for representing the density of the Mg$^+$ ions around
individual galaxies.  The expected Mg\,II absorption equivalent width,
$\bar{W}$ as a function of galaxy impact parameter can be calculated
following
\begin{equation}
\bar{W}(2796)=\frac{\bar{W}_0}{\sqrt{\rho^2+a_h^2}}\tan^{-1}{\sqrt{\frac{R_{\rm gas}^2-\rho^2}{\rho^2+a_h^2}}}
\end{equation}
at $\rho\le R_{\rm gas}$ and $\bar{W}(2796)=0$ otherwise.  The core
radius $a_h$ is defined to be $a_h=0.2\,R_{\rm gas}$ and does not
affect the expected $\bar{W}(2796)$ at large $\rho$.  The extent of
Mg\,II absorbing gas scales with the luminosity of the absorbing
galaxy according to
\begin{equation}
\frac{R_{\rm gas}}{R_{{\rm gas}*}}=\left(\frac{L_B}{L_{B_*}}\right)^{\beta}.
\end{equation}
Following the definition of Equation (2), $R_{{\rm gas}*}$
characterizes the gaseous extent of $L_*$ galaxies.  The gaseous
extent of fainter/more luminous galaxies differs from $R_{{\rm gas}*}$
according the scaling relation.

  To determine the values of $\bar{W}_0$, $R_{{\rm gas}*}$, and
$\beta$ that best describe the data, we perform a maximum-likelihood
analysis that includes the upper limits in our galaxy and Mg\,II
absorber pair sample.  The likelihood function of this analysis is
defined as
\begin{eqnarray}
{\cal L} & = & \left( \prod_{i=1}^{n} \exp \left\{ -\frac{1}{2} \left[ \frac{y_i -
\bar{y}(\rho_i, L_{B_i})}{\sigma_i} \right]^2 \right\} \right)\times \\ \nonumber 
 & & 
\left( \prod_{i=1}^m \int_{y_i}^{-\infty} dy' \exp \left\{ -\frac{1}{2} \left[ 
\frac{y' - \bar{y}(\rho_{i}, L_{B_i})}{\sigma_i} \right]^2 \right\} \right),
\end{eqnarray}
where $y_i=\log\,W_i$, $\bar{y}=\log\,\bar{W}$, and $\sigma_i$ is the
measurement uncertainty of $y_i$.  The first product of Equation (3)
extends over the $n$ measurements and the second product extends over
the $m$ upper limits.  (This definition of the likelihood function is
appropriate if the residuals about the mean relationship are normally
distributed.)  Because $\sigma_i$ may include significant intrinsic
scatter (which presumably arises due to intrinsic variations between
individual galaxies) as well as measurement error, $\sigma_i$ is taken
to be the quadratic sum of the cosmic scatter $\sigma_c$ and the
measurement error $\sigma_{m_i}$
\begin{equation}
\sigma_i^2 = \sigma_c^2 + \sigma_{m_i}^2,
\end{equation}
where the intrinsic scatter is defined by
\begin{equation}
\sigma_c^2 =  {\rm med} \left( \left\{ y_i - \bar{y}(\rho_{i},L_{B_i}) -
\frac{1}{n} \sum_{j=1}^n \left[ y_j - \bar{y}(\rho_{j},L_{B_j}) \right]
\right\}^2 - \sigma_{m_i}^2 \right).
\end{equation}
Because $\sigma_c$ depends on the maximum-likelihood solution
$\bar{y}=\log\,\bar{W}(\rho_{i},L_{B_i})$, the maximum-likelihood
solution is obtained iteratively with respect to equations (1) and
(2).  The maximum-likelihood analysis yields
\begin{equation}
\beta=0.35\pm 0.05,
\end{equation}
and
\begin{equation}
R_{{\rm gas}*}=91_{-8}^{+3}\ h^{-1}\ {\rm kpc},
\end{equation}
and $\bar{y}_0=\log\,\bar{W}_0=-0.25\pm 0.05$ for the best-fit $R_{\rm
gas*}$.  The errors represent the 1-$\sigma$ uncertainties of the
best-fit parameters.  In addition, we estimate a cosmic scatter of
$\sigma_c=0.25$ dex.  The best-fit isothermal model is presented in
Figure 4 (solid curve).

\begin{figure}
\begin{center}
\includegraphics[scale=0.45]{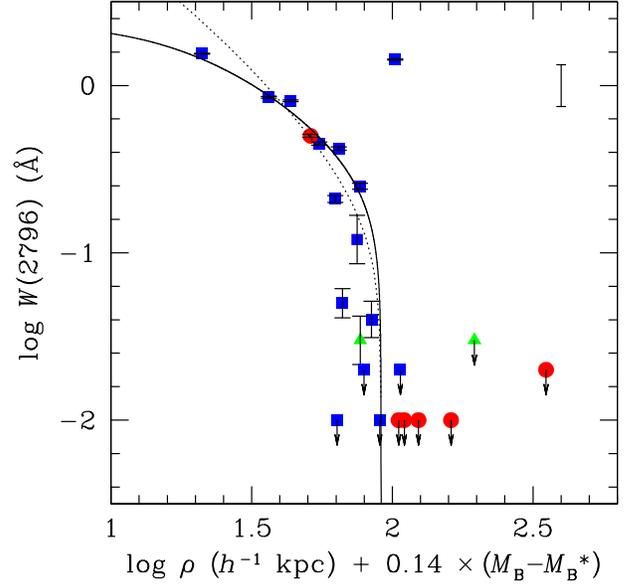}
\caption{Comparison of the rest-frame absorption equivalent width
$W(2796)$ versus galaxy impact parameter $\rho$ scaled by galaxy
$B$-band luminosity.  Symbols are the same as those in Figure 3.  The
solid curve represents the best-fit model based on an isothermal
density profile of the Mg$^+$ ions, while the dotted curve shows the
expectation from an NFW density profile.  The errorbar in the
upper-right corner represents the intrinsic scatter of the best-fit
model, in comparison to the observations, as defined in Equation (5).}
\end{center} 
\end{figure}

  The results of our maximum likelihood analysis indicate that the
extent of Mg\,II gaseous halo scales with galaxy $B$-band luminosity
according to $(L_B/L_{B_*})^\beta$ and $\beta=0.35\pm 0.05$, and that
typical $L_*$ galaxies possess an extended Mg\,II halo of radius
$R_{\rm gas}\approx 91\ h^{-1}$ kpc.  It is worth noting that the
best-fit $91_{-8}^{+3}\ h^{-1}$ kpc characterizes the gaseous extent
of $L_*$ galaxies, beyond which no Mg$^+$ ions are found to the limit
of $W_r(2796)=0.01$ \AA.  Adopting the best-fit scaling relation, the
gaseous extent of $0.1\,L_*$ galaxies is expected to be $\approx 41 \
h^{-1}$ kpc.  Figure 4 shows the relation of $W_r(2796)$ vs.\
luminosity scaled $\rho$, in which seven galaxies at $> 100\ h^{-1}$
kpc do not have a corresponding Mg\,II absorption feature to the limit
of $W(2796)=0.01$ Ang.  These non-detections provide the strongest
constraints on $R_{{\rm gas}*}$.

  To examine possible bias in $R_{{\rm gas}*}$ as a result of the
initial sample definition that consider only galaxies at $\rho \apll
r_0=100\ h^{-1}$ kpc, we repeat the likelihood analysis adopting
$r_0=80\ h^{-1}$ kpc and find that the best-fit results remain the
same.  Choosing a smaller $r_0$ reduces the sample substantially and
the parameters become poorly constrained.  Furthermore, we note that
the results are also insensitive to the choice of model gaseous
profile.  We obtain similar results, adopting a Navarro-Frenk-White
(NFW) profile commonly used for describing the profiles of dark matter
halos (Navarro \etal\ 1997).  The best-fit NFW profile is also
presented in Figure 4 (dotted curve).  Comparisons of observations and
different model profiles show that an isothermal model profile of the
underlying gaseous halo provides a good fit to the observed $W(2796)$
vs.\ $\rho$ relation.

  In summary, Figure 4 demonstrates that accounting for the scaling
relation of gaseous extent with galaxy luminosity, the scatter in the
$W(2796)$ versus $\rho$ relation is substantially reduced.
Furthermore, the $W(2796)$ versus $\rho$ relation scaled by galaxy
intrinsic luminosity allows us to estimate a mean covering factor of
Mg$^+$ ions average across galaxies of a wide range of luminosity.  We
find $\langle\kappa\rangle=80-86$ \% within $R_{\rm gas}$, including
the uncertainty in $R_{{\rm gas}*}$ from Equation (7).

\section{DISCUSSION}

  Using the sample of 13 galaxy--Mg\,II absorber pairs and 10 galaxies
at $\rho\apll 100\ h^{-1}$ kpc that do not give rise to Mg\,II
absorption to a sensitive upper limit, we find that galaxies at larger
impact parameters produce on average weaker Mg\,II absorbers.  This
anti-correlation is substantially improved, when accounting for the
intrinsic luminosity of individual galaxies.  In addition, there
exists a distinct boundary at $\rho=R_{\rm gas}$, beyond which no
Mg\,II absorbers with $W(2796)\ge 0.01$ \AA\ are found.  A maximum
likelihood analysis shows that the observations are best described by
an isothermal density profile of the gas and a scaling relation
$R_{\rm gas}=91_{-8}^{+3}\times (L_B/L_{B_*})^{(0.35\pm 0.05)}\
h^{-1}$ kpc with a mean covering factor of
$\langle\kappa\rangle=80-86$ \% at $\rho\le R_{\rm gas}$.  The
best-fit profile applies to galaxies of $0.03\,L_{B_*}$ to
$1.6\,L_{B_*}$ and therefore predominantly cold-mode halos.  For
higher mass halos, we expect from TC08 that the corresponding
absorption strength for a given $\rho$ is reduced, namely a smaller
$\bar{W}_0$ in Equation (1).  In this section, we discuss the
implications drawn from the results of our analysis.

\subsection{Comparisons with Previous Studies}

  Adopting the best-fit scaling relation and the gaseous profile
described in Equation (1), we derive a corresponding halo size of
$\rho\approx 69\ h^{-1}$ physical kpc for galaxies of
$M_B-5\,log\,h=-19.8$ at $W(2796)=0.3$ \AA.  This halo size is
consistent with the finding of Lanzetta \& Bowen (1990) and the
expectation inferred from known Mg\,II frequency function (Nestor
\etal\ 2005; see the descriptions in \S\ 5.2), but is larger than what
was reported in Steidel (1995).  Steidel (1995) reported a
characteristic size of Mg\,II halo of $R_*\approx 35\ h^{-1}$ kpc for
$L_*$ (corresponding to $M_{B_*}-5\,\log\,h=-19.7$ for the cosmology
adopted in this paper) galaxies, but it is not clear at what $W(2796)$
limit the halo size was estimated.  At $\rho=35\ h^{-1}$ kpc, we
expect $\bar{W}(2794)=0.9$ \AA\ from the best-fit model.

  The high gas covering factor derived from our analysis is consistent
with what is inferred from the sample of Kacprzak \etal\ (2008).
Applying the best-fit luminosity scaling relation of gaseous extent,
we find that four of the 28 galaxies within the expected gaseous
extent for $W_r(2796)> 0.3$ \AA\ absorbers in the Kacprzak \etal\
sample have a corresponding Mg\,II weaker than the 0.3 \AA\ threshold.
The inferred covering fraction is $\approx 86$\%.  At the same time,
this high covering fraction disagrees with the value reported in Tripp
\& Bowen (2005), the only other on-going survey program designed to
constrain gas covering fraction by searching for corresponding Mg\,II
absorbers at the locations of known {\it field} galaxies\footnote{We
note two earlier surveys by Bechtold \& Ellingson (1992) and Bowen
\etal\ (1995) that include a dominant fraction of galaxies in dense
cluster environment.  Both studies reported a substantially lower
covering fraction, $\langle\kappa\rangle\apll 20$ \% for $W(2796)>0.1$
absorbers.  The low covering fraction of strong absorbers is
understood, because galaxies clusters reside in massive dark matter
halos of $M_h> 10^{14}\,\hmsol$ and the fraction of ``cold'' gas in
these massive halos is expected to be reduced (see TC08).}.  The
authors obtained follow-up spectra of background QSOs that are located
within $\rho=10-39\ h^{-1}$ physical kpc of 20 known galaxies at
$z_{\rm gal}=0.31-0.55$.  They found that 10 of the 20 galaxies do not
produce Mg\,II absorbers to within a 2-$\sigma$ upper limit of
$W(2796)=0.1$ \AA, and concluded that the incidence of Mg$^+$ ions is
$\approx\,50$ \% at $\rho\apll 40\ h^{-1}$ kpc in individual galactic
halos.

  Because more luminous galaxies have larger extended Mg\,II halos
according to the scaling relation presented in Equations (2), (6), and
(7), this discrepancy can be understood if many of the galaxies in
Tripp \& Bowen are located at impact parameters that are larger than
their expected $R_{\rm gas}$.  Recall that we obtained a similar
estimate of $\langle\kappa\rangle\approx 57$ \% at $\rho< 100\ h^{-1}$
kpc (or $\langle\kappa\rangle\approx 50$ \% at $\rho< 45\ h^{-1}$
kpc), {\it prior to} accounting for the scaling relation between
$R_{\rm gas}$ and $L_B$ in individual galaxies (\S\ 4).  

  In addition, the sample in Tripp \& Bowen includes galaxies of
luminosity ranging from $0.3\,L_*$ to $5\,L_*$.  Those galaxies that
are more luminous than $L_*$ are also expected to reside in more
massive halos.  At $z=0$, $5\,L_*$ galaxies are expected to reside in
$> 10^{13}\,\hmsol$ halos (Tinker \etal\ 2007).  If the host dark
matter halos are substantially more massive than $10^{12.0}\,\hmsol$,
then we expect from the halo occupation analysis in TC08 that they do
not contribute significantly to strong Mg\,II absorbers.  In summary,
further inspections of the properties of these galaxies and estimates
of the gas covering fraction $\kappa$ as a function of radius and
$W(2796)$ are crucial for resolving this discrepancy.

\subsection{The Origin of Mg\,II Absorbers}

  The scaling relation and the constraint of $\langle\kappa\rangle$
together also allow us to test whether the Mg\,II absorbers trace a
significant and representative portion of the galaxy population or
merely a special class such as those galaxies undergoing a starburst
episode.  If extended Mg\,II halos are a generic feature of
field\footnote{We define field galaxies as those that are not
associated with a known cluster.} galaxies at all epochs, then we can
derive the expected number density of Mg\,II absorbers by combining
known space density of galaxies (versus luminosity) and the product of
the cross section and covering factor of the Mg$^+$ ions.  An
agreement with observational results from surveys of Mg\,II absorbers
would challenge the scenario that attributes a large fraction of these
absorbers to starburst systems\footnote{However, we will not to able
to rule out a scenario in which the blow-out winds are common in most
galaxy histories.  Given enough time, long-lived outflowing gas gets
mixed in with existing halo gas and newly accreted material from the
IGM.  There is little distinction at this point between outflows and
accreted clouds.  It is also not clear whether the outflow material
would maintain the same degree of complex gas kinematics as those that
are closer to the parent starburst regions.}.

  The predicted number density of Mg\,II absorption systems arising in
the extended gaseous halos of galaxies may be evaluated according to
\begin{eqnarray}
\frac{d\,{\cal N}(W\ge W_{lim})}{d\,z} & = & \frac{c}{H_0} \frac{(1 + z)^2}{\sqrt{\Omega_{\rm M}\,(1+z)^3 + \Omega_{\rm \Lambda}}} \\
& &  \times \int_0^\infty
d\left(\frac{L_B}{L_{B_*}}\right)\,\Phi(L_B,z)\,\sigma(L_B,W_{lim})\,\kappa, \nonumber
\end{eqnarray}
where $c$ is the speed of light, $\Phi(L_B,z)$ is the galaxy
luminosity function, $\sigma$ is the gas cross section for producing
Mg\,II absorbers of $W(2796)\ge W_{lim}$ that scales with galaxy $B$-band
luminosity, and $\kappa$ is the halo covering factor.  Substituting
the scaling relationship according to Equations (2), (6), and (7), we
find
\begin{eqnarray}
\frac{d\,{\cal N}(W\ge W_{lim})}{d\,z} & = & \frac{c}{H_0}\frac{\pi\,\langle\kappa\rangle\,[R'_{{\rm gas}*}(W_{lim})]^2\,(1 + z)^2}{\sqrt{\Omega_{\rm M}\,(1+z)^3 + \Omega_{\rm \Lambda}}} \\
& & \times\int_0^\infty d\left(\frac{L_B}{L_{B_*}}\right)\,\left(\frac{L_B}{L_{B_*}}\right)^{2\,\beta}\,\Phi(L_B,z)\,, \nonumber
\end{eqnarray}
where $R'_{{\rm gas}*}(W_{lim})$ is the extent of Mg\,II halos that
corresponds to $W(2796)=W_{lim}$.

  To evaluate Equation (9), we adopt the luminosity function of the
blue galaxy population at $z_{\rm gal}=0.3-0.5$ from Faber \etal\
(2007), which is characterized by $M_{B_*}-5\,\log=-19.8$,
$\phi_*=9.6\times 10^{-3}\ h^{3}$ Mpc$^{-3}$, and a faint-end slope
$\alpha=-1.3$.  Including galaxies of $L_B\ge 0.01\,L_{B_*}$ and our
best-fit $R_{{\rm gas}*}$ and $\langle\kappa\rangle$, Equation (9)
yields $d\,{\cal N}/dz=0.028-0.03$ for $W(2796)\ge 2$ \AA\ absorbers,
$d\,{\cal N}/dz=0.21-0.23$ for $W(2796)\ge 1$ \AA\ absorbers, and
$d\,{\cal N}/dz=0.88-1.07$ for $W(2796)\ge 0.3$ \AA\ absorbers at
$z=0.5$.  The predictions based on galaxies more luminous than
$0.01\,L_*$ agree well with the observed $d\,{\cal N}/d\,z$ from
Nestor \etal\ (2005), lending strong support for the hypothesis that
the Mg\,II absorbers trace a significant and representative portion,
rather than a sub-class, of the galaxy population.

  The model profile of Equation (1) has been proved to be a good
representation of the galaxy and absorber pairs at $\rho=16.3-100\
h^{-1}$ kpc, when the scaling relation of Equations (2), (6), and (7)
is accounted for.  An important test is to examine whether the model
remains a good fit at smaller $\rho$ and strong absorbers.
Identifying galaxy--absorber pairs at small impact parameters is
difficult, because the glare of the background QSO often prohibits
identification of faint galaxies at angular distances $\theta<2''$
that corresponds to $\sim 10\ h^{-1}$ physical kpc at $z\apll 0.5$.

  Recent studies of intervening absorption-line systems along the
lines of sight toward the optical afterglows of distant $\gamma$-ray
bursts (GRB) have uncovered a number of strong Mg\,II absorbers (e.g.\
Prochter \etal\ 2006b).  Unlike QSOs, GRB afterglows disappear after a
while, permitting exhaustive searches of faint galaxies at
$\theta<2''$ that give rise to strong Mg\,II absorbers along the lines
of sight.  A redshift survey of galaxies along the sightline toward
GRB\,060418 ($z_{\rm GRB}=1.491$) has uncovered the absorbing galaxies
for three strong Mg\,II absorbers at $z=0.603-1.107$ (Pollack \etal\
2008, in preparation), similar to the redshift range covered by our
sample.  The rest-frame $B$-band magnitudes of the galaxies range from
$M_B-5\,\log\,h=-16.8$ to $M_B-5\,\log\,h=-18.3$; Impact parameters of
the galaxies range from $\rho=7.5\ h^{-1}$ kpc to $\rho=16.5\ h^{-1}$
kpc.  Adopting the best-fit scaling relation, we include the three
galaxy--Mg\,II absorber pairs in the $W(2796)$ versus $\rho$ plot for
comparison (star points in Figure 5).  The good agreement between the
best-fit model and the absorbers found along GRB sightlines provides
further support for our model at small impact parameters.

\begin{figure}
\includegraphics[scale=0.45]{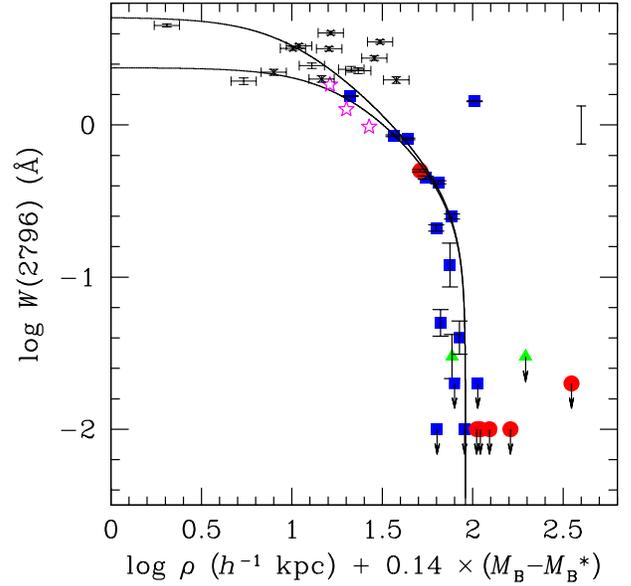}
\caption{The $W(2796)$ versus $\rho$ correlation scaled by galaxy
$B$-band luminosity presented in Figure 4, but including pairs from
Pollack \etal\ (2008; star points) and Bouch\'e \etal\ (2007; points
with horizontal errorbars) for comparisons.  The solid curves
represent the best-fit model based on an isothermal density profile,
with $a_h=0.1\,R_{\rm gas}$ (top curve) and $a_h=0.2\,R_{\rm gas}$
(bottom curve).  The pairs of Pollack \etal\ are for galaxies and
Mg\,II absorbers identified at $z=0.6-1$ along the sightline toward
GRB\,060418.  The errorbars in these data points are smaller than the
size of the symbols.  The pairs of Bouch\'e \etal\ are galaxies found
at the positions of known Mg\,II absorbers based on the presence of
H$\alpha$ emission lines.  Errorbars represent the uncertainties of
the inferred $M_B$ from the $M_B$ versus $L(\ha)$ correlation
published in Tresse \etal\ (2002).  Crosses represent the absorbers in
the Bouch\'e sample that are likely ($\apg 40$ \% probability) to be a
damped \lya\ absorber of $N(\hI)\ge 2\times 10^{20}$ \cmjj\ based on
the presence of strong Fe\,II and Mg\,I transitions (Rao \etal\
2006).}
\end{figure}

\subsection{Are Strong [$W(2796)>1$ \AA] Absorbers at $z\sim 1$ Produced Primarily in Starburst Outflows?}

  As described in \S\ 1, the large line width of $W(2796)>1$ \AA\
Mg\,II absorbers has recently been interpreted as due to outflow
motions in starburst driven winds.  This scenario was initially
motivated by the observed inverse correlation between $W(2796)$ and
the clustering amplitude of Mg\,II absorbers at $\langle z\rangle=0.6$
(Bouch\'e \etal\ 2006).  In TC08, we have shown that this inverse
correlation can be understood as due to an elevated clustering
amplitude of $W(2796)\apll 1$ \AA\ absorbers from contributions of
survived cold clouds in massive hot halos, rather than a suppressed
clustering strength of $W(2796)\apg 2$ \AA\ absorbers.  In addition,
we have shown in \S\S\ 4, 5.1, and 5.2 that extended Mg\,II halos are
a generic feature of field galaxies over a wide luminosity range.  The
predicted number density of the absorbers from adopting the scaling
relation, Equations (2), (6), and (7), and the galaxy luminosity
function agrees well with observations.  Specifically, we expect
$d\,{\cal N}/dz\approx 0.03$ for $W(2796)\ge 2$ \AA\ absorbers, in
comparison to the observed $d\,{\cal N}/dz=0.03-0.04$ for $W(2796)\ge
2$ \AA\ absorbers at $z\approx 0.5$ from Nestor \etal\ (2005).  While
there is no clear indicator one can apply to unambiguously distinguish
between gas inflows and outflows in the distant universe, we discuss
several caveats related to the starburst outflow scenario.

  If the absorbers are produced in starburst driven outflows, then a
large fraction of the absorbing galaxy population might be expected to
exhibit disturbed morphology (Mobasher \etal\ 2004).  The Mg\,II
absorbing galaxies in our sample exhibit regular disk morphologies in
the HST images shown in Figure 2.  In particular, the strong absorber
found with $W(2796)=1.55$ \AA\ at $z=0.892$ toward 3C336 is associated
with an edge-on disk galaxy at $\rho= 16.3\ h^{-1}$ kpc.  The three
Mg\,II absorbers of $W(2796)>1$ \AA\ found toward GRB\,060418 also
exhibit regular disk structures in high-resolution images obtained
both in space using HST and on the ground using LGAO on the Keck
telescopes (Pollack \etal\ 2008, in preparation).  Kacprzak \etal\
(2007) also showed that there is little correlation between
$W_r(2796)$ and galaxy asymmetry for Mg\,II absorbers of
$W_r(2796)>1.4$ \AA.  While some strong Mg\,II absorbers have been
presented to show evidence of arising in expanding superbubble shells
(Bond \etal\ 2001), available high-resolution images of other strong
Mg\,II absorbers do not support outflow being principally responsible
for the observed strong absorbers.  This conclusion is also consistent
with the spectral properties of Mg\,II-selected galaxies published by
Guillemin \& Bergeron (1997)

  On the other hand, Bouch\'e \etal\ (2007) presented new observations
from a targeted search for corresponding H$\alpha$ emission in the
vicinity of known Mg\,II absorbers.  These authors successfully
identified H$\alpha$ emission near 14 of 21 Mg\,II absorbers with
$W(2796)>1.9$ \AA\ at $z\sim 1$.  The integrated raw\footnote{No
extinction correction is applied to $L(\ha)$ in our analysis, because
the dust content is expected to vary substantially in $z\sim 1$
galaxies (e.g.\ Tresse \etal\ 2002) and is unknown for galaxies in the
Bouch\'e \etal\ sample.  Bouch\'e \etal\ (2007) applied a constant
correction to all observed \ha\ flux in their calculation, resulting
in a simple scaling of the observed values.  The scatter in their
measurements induced by dust remains the same.  For consistency, we
consider in the following discussion only observed values without
extinction correction both for the absorbing galaxies and for
comparison results, such as the \ha\ luminosity function and the
$M_B-L(\ha)$ correlation, from the literature.} \ha\ luminosities
range from $L_{\ha}=6.8\times 10^{40}\ h^{-2}$ erg s$^{-1}$ to
$L_{\ha}=8.4\times 10^{41}\ h^{-2}$ erg s$^{-1}$.  The impact
parameters range from $\rho=1.4\ h^{-1}$ kpc to $\rho=35\ h^{-1}$ kpc.
Adopting a mean extinction correction $A_V=0.8$ mag, the authors
inferred a mean star formation rate of $\langle{\rm SFR}\rangle=5.9\
h^{-2}$ M$_\odot$ yr$^{-1}$ for these galaxies.

  The \ha\ luminosity function at $z\sim 1$ has been studied by a
number of groups (see Doherty \etal\ 2006 for a list of references).
It is characterized by a Schechter function of $L_{{\ha}_*}=5\times
10^{41}\ h^{2}$ erg s$^{-1}$, $\phi_*=0.013\ h^3\,{\rm Mpc}^{-3} $,
and $\alpha=-1.3$.  The galaxies detected in \ha\ by Bouch\'e \etal\
correspond to $0.14-1.68\,L_{\ha_*}$ at $z\sim 1$, and therefore have
only modest $L(\ha)$.  Integrating the \ha\ luminosity function, we
obtain a mean space density of 0.034 $h^{3}$ Mpc$^{-3}$ for galaxies
of brighter than $0.1\,L_{{\ha}_*}$.  This is already 50\% of all
galaxies more luminous than $0.01\,L_{B_*}$ at $z\sim 1$, according to
the best-fit luminosity function of the blue galaxy population from
Faber \etal\ (2007).  This exercise shows that \ha\ emission is
commonly seen in $z\sim 1$ galaxies and the presence of modest
$L(\ha)$ does not argue for the presence of starburst outflows, unless
the outflow material is long-lived.

  For comparison, we include the 14 galaxy and absorber pairs from
Bouch\'e \etal\ (2007) in the $W(2796)$ versus $\rho$ plot in Figure
5, applying the same scaling relation described in Equations (2), (6),
and (7).  We estimate $M_B$ of these galaxies based on the published
$M_B-L(\ha)$ correlation in Tresse \etal\ (2002).  We adopt the
observed $L(\ha)$ and the scaling relation
$M_B=73.5-2.27\times\log\,L_(\ha)$ that best describes the data in
Tresse \etal\ (2002).  At small impact parameters, the model
predictions depend sensitively on the choice of $a_h$ in Equation (1).
In addition to the fiducial model, we also include in the figure a
model with $a_h=0.1\,R_{\rm gas}$.

  Nine of the 14 Mg\,II absorbers are considered likely ($>40$ \%)
damped \lya\ absorbers (DLAs) of $N(\hI)\ge 2\times 10^{20}$ \cmjj\
based on the presence of strong Fe\,II and Mg\,I transitions (Rao
\etal\ 2006).  DLAs are the high-redshift analog of neutral gas
regions that resembles the disks of nearby luminous galaxies (e.g.\
Wolfe \etal\ 2005).  We therefore expect that the impact parameter
distribution of galaxy-DLA pairs represents the typical extent of H\,I
disks at high redshifts.  We highlight the potential DLAs in the
Bouch\'e \etal\ sample as crosses in Figure 5.  After applying the
best-fit scaling relation for Mg\,II absorbers, we find that the
impact parameters of these galaxy and potential DLA pairs range from
$8\ h^{-1}$ kpc to as high as $38\ h^{-1}$ kpc.  The three largest
separation pairs have $\rho'>28\ h^{-1}$ kpc, where $\rho'$ is the
luminosity scaled projected distance.  The scaling of impact parameter
is justified, because smaller/fainter galaxies have on average smaller
\hI\ disks (e.g.\ Cayatte \etal\ 1994).  The large impact parameters,
together with the likely high $N(\hI)$, suggest that the galaxies
responsible for these absorbers (presumably at smaller angular
distances from the QSOs) may still be missing.  For the remaining
sample, our model is in reasonably good agreement with their data.

  In summary, we find a lack of empirical measurements that argue for
starburst outflows to be a dominant mechanism for producing the
observed strong Mg\,II absorbers at $z\sim 1$.  On the other hand,
both the observed number density of Mg\,II absorbers and the $W(2796)$
versus $\rho$ correlation can be explained by extended Mg\,II halos
around typical field galaxies.  The fraction of Mg\,II absorbers
originating in starburst outflows is therefore expected to be small at
$z\sim 1$, unless starburst outflows are a common feature in field
galaxies.  We caution, however, that the same conclusion cannot be
automatically applied to absorbers at $z\sim 2$ before a similar
analysis is performed.

\subsection{Constraints on the Extended Gaseous Halos Around Galaxies}

  The sharp decline of $W(2976)$ found at $\rho\approx 90\ h^{-1}$ kpc
in Figure 4 clearly indicates that Mg$^+$ ions have a finite extent in
the extended gaseous halos around galaxies.  This is reminiscent of
the sharp boundary between the C\,IV absorbing and non-absorbing
regions around $\langle z\rangle=0.4$ galaxies reported in Chen \etal\
(2001b).

  Given the nature of a photo-ionized gas (Bergeron \& Stas\'inska
1986; Hamann 1997), we examine whether the finite extent of Mg$^+$
ions can be understood as due to photo-ionization of the absorbing
clouds at large radii.  In a two-phase halo model (e.g.\ Mo \&
Miralda-Escud\'e 1996), cold clouds that are responsible for producing
the absorption features are pressure confined in a hot medium.  As the
total gas density declines toward larger radii, the density of the
cold clouds also decreases and the clouds become optically thin to the
ultraviolet photons that can quickly ionize the Mg$^+$ ions
(I.P.\,$=$\,15.035 eV).  With a known background radiation field, the
location where the ionization transition occurs may be adopted to
constrain the density of cold gas at large radii.

  We carry out the photo-ionization analysis, using the Cloudy
software (Ferland \etal\ 1998; version 06.02), and calculate the
ionization fractions of Mg$^+$ ions for a grid of models with
different total gas density and metallicity.  We adopt the spectral
shape of the UV background radiation field of Haardt \& Madau in
Cloudy for the incident ionizing flux, and scale the radiation
intensity to the background UV radiation field inferred from QSO
proximity effect (Scott \etal\ 2002).  We adopt a plane parallel
geometry for the clouds with a thickness of 100 pc, which is motivated
by the observed coherent length of Mg\,II absorbers toward lensed QSOs
(Rauch \etal\ 2002).  The slab of gas is illuminated by the ionization
radiation field on both sides.

\begin{figure}
\begin{center}
\includegraphics[scale=0.4]{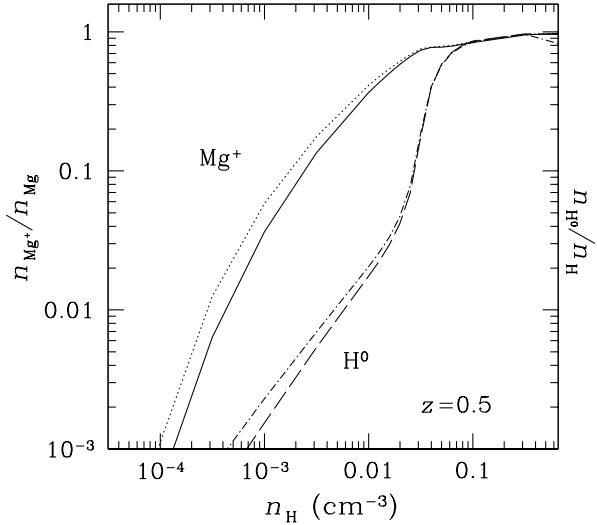}
\caption{Ionization fractions of Mg$^+$ ions and H$^0$ atoms versus
total gas density $n_{\rm H}$ for photo-ionized gas.  The values are
calculated using the Cloudy software (Ferland \etal\ 1998; version
06.02) for gaseous clouds of plane parallel geometry and 100 pc
thickness.  The input ionizing radiation intensity is set to be
$J_\nu(912\AA)=7\times 10^{-23}$ erg s$^{-1}$ Hz$^{-1}$ cm$^{-2}$
Sr$^{-1}$ suitable for the $z<1$ universe (Scott \etal\ 2002).  We
perform the calculations, assuming solar (dotted curve for Mg and
dot-dashed curve for H) and $1/10$ solar metallicity (solid curve for
Mg and dashed curve for H).}
\end{center} 
\end{figure}

  Figure 6 shows the expected fractions of Mg$^+$ ions (solid and
dotted curves) as a function of the total gas density $n_{\rm H}$.  We
include the calculated fraction of H$^0$ atoms (dashed and dash-dotted
curves) for comparison.  The solid and dashed curves are for gas of
$1/10$ solar metallicity, while the dotted and dash-dotted curves are
for gas of solar metallicity.  Our photo-ionization calculations
indicate that indeed the fraction of Mg$^+$ declines linearly at
$n_{\rm H}\apll 0.025\ {\rm cm}^{-3}$ for a mean radiation intensity of
$J_\nu(912\AA)=7\times 10^{-23}$ erg s$^{-1}$ Hz$^{-1}$ cm$^{-2}$
Sr$^{-1}$ (which is appropriate for the observed background radiation
field at $z<1$ Scott \etal\ 2002).  However, the transition is not as
sharp as what is seen for hydrogen atoms.  Including the estimated
ionization fraction from the Cloudy models and adopting the density
profile parameterized in Equation (1), we derive the expected
correlation between $W_r(2796)$ and $\rho$ and find that
photo-ionization is insufficient for explaining the presence of the
sharp decline in the observed Mg\,II absorption strength.

  This extent of Mg\,II gaseous halos ($\approx 90\ h^{-1}$ kpc around
typical $L_*$ galaxies) is remarkably similar to the extent of C\,IV
gaseous halo ($\approx 110\ h^{-1}$ kpc around typical $L_*$ galaxies,
after correcting for the $\Lambda$-cosmology) reported by Chen \etal\
(2001b).  On the basis of 14 galaxy and C\,IV absorber pairs and 36
galaxies that do not produce corresponding C\,IV absorption lines to
within sensitive upper limits, these authors find that the extent of
C\,IV-absorbing gas around galaxies scales with galaxy $B$-band
luminosity as $R_{\rm gas}({\rm C\,IV}) \propto 110 \times L_B^{0.5
\pm 0.1}\ h^{-1}$ kpc (corrected for the $\Lambda$ cosmology).  In
addition, there exists a sharp boundary between C\,IV absorbing and
non-absorbing regions at $R_{\rm gas}({\rm C\,IV})$.  But the scatter
in the observed C\,IV absorption strengths at $\rho< R_{\rm gas}({\rm
C\,IV})$ between different galaxies appears to be significantly larger
than what is seen in Figure 4, and there is a lack of anti-correlation
between the observed C\,VI absorbing strength and galaxy impact
parameter.

  Given the distinct ionization potentials of the C$^{3+}$ and Mg$^+$
ions, {\it the location where the abrupt transition between absorbing
and non-absorbing regions occurs can be understood as a critical
radius below which cool clouds can form and stablize in an otherwise
hot halo}.  This two-phase model to interpret QSO absorption line
systems was formulated in Mo \& Miralda-Escud\'e (1996) and later
re-visited by Maller \& Bullock (2004).

  Taking into account the appropriate photo-ionization condition of
the cold clouds, Mo \& Miralda-Escud\'e presented the expected Mg\,II
and C\,IV absorption strength versus radius for halos of different
mass.  Their models showed a lack of radial dependence on the C\,IV
absorption strength and a tight correlation between Mg\,II $W_r(2796)$
and $\rho$ for halos of a wide range of mass.  The differences in the
radial dependence of different ions is understood as being due to the
photo-ionization condition of cold clouds in the UV background
radiation field.  These model expectations agree with observational
findings very well.  

  The agreement between observations and a simple two-phase model is
encouraging, although it is not clear how the two-phase model applies
to cold-mode halos (e.g.\ Dekel \& Birnboim 2006).  Measurements of
the relative abundances between C$^{3+}$ and Mg$^+$ as a function of
radius will provide additional support for the origin of these
metal-line absorbers, further constraining the metallicity and density
of halo gas around galaxies.

\subsection{Implications for the Halo Occupation Analysis and Future Work}

Our initial halo occupation analysis (TC08) included two principal
assumptions for the extent of Mg$^+$.  First, we assumed that the
extent of the cold gaseous halo is $R_{\rm gas,12}=50\,h^{-1}$
physical kpc for $M_h=10^{12}\,\hmsol$ halos, which is roughly $1/3$
of the halo radius $R_{200}$ at $z=0.6$.  Recall that the standard
definition of a dark matter halo is an object with a mean interior
density of 200 times the background density.  Second, the gaseous
extent scaled with the halo mass as $M_h^{t}$ and $t=1/3$, following
the expectation of the scaling relation between halo mass and the
corresponding virial radius.  The results presented in \S\ 4 offer a
direct test of these assumptions.

To compare the empirical constraints of Equations (6) \& (7) with our
initial model assumptions, we first derive the corresponding dark
matter halo mass $M_h$ and their halo radius $R_{200}$ for galaxies of
known absolute $B$-band magnitudes.  The adopted $M_{\rm B_*}$ for the
analysis in \S\ 4 corresponds to a magnitude of $M_{\rm
b_J}-5\,\log\,h=-19.7$ in the bandpass of the Two-Degree Field Galaxy
Redshift Survey (2dFGRS, Colless \etal\ 2001).  The 2dFGRS probes
galaxies at $z\sim 0.1$.  Excluding galaxies that exist as satellites
in a cluster environment\footnote{We note that the fraction of
satellite galaxies is $\lesssim 15$ \% for $L_*$ galaxies (Tinker
\etal\ 2007; ZCZ07).  In addition, none of the galaxies in our sample
are observed to be part of a larger group or cluster.  Here we carry
out our calculations, assuming that most of galaxies in our sample are
central galaxies in their dark matter halos.}, the mean halo mass for
$M_{\rm B_*}$ galaxies is $10^{12.5}\,h^{-1}\,\hmsol$ at $z\sim 0.1$
(Tinker \etal\ 2007; van den Bosch \etal\ 2007).  At $z\sim 1$,
galaxies of this magnitude on average reside in $10^{11.9}\,\hmsol$
halos (ZCZ07).  Interpolating in $\log\,(1+z)$, we find that galaxies
of $M_{\rm B_*}-5\,\log\,h=-19.8$ are expected to reside in halos of
$M_h=10^{12.3}\,\hmsol$ at $z=0.4$ and that the halo radius for
$M_{\rm B_*}$ galaxies is $R_{200}=212\,h^{-1}$ physical kpc.  The
best-fit $R_{\rm gas*}$ from Equation (7) therefore implies $R_{\rm
gas}\approx 0.4\,R_{200}$.

Next, we derive the expected scaling relation between $R_{\rm gas}$
and halo mass $M_h$.  The halo occupation analysis of 2dFGRS galaxies
from Tinker \etal\ (2007) showed that the relationship between $M_h$
and luminosity for galaxies of $L_{b_J}\lesssim L_{*}$ at $z\sim 0.1$
is $M_h = 10^{12.5}\,(L_B/L_{B_*})^{1.3}\,\hmsol$.  this monotonic
relationship is appropriate for galaxies that reside at the {\it
centers} of their dark matter halos and are the brightest galaxy in
the halo.  A similar scaling relation was obtained from DEEP2 data by
ZCZ07 for galaxies at $z\sim 1$.  Because there is a
monotonic relationship between halo mass and galaxy luminosity, the
scaling relation of $R_{\rm gas}$ with $L_B$ in Equation (2) is
equivalent to a scaling relation between $M_h$ and $R_{200}$ that are
related by $M_h\propto R_{200}^3$.  Adopting $\beta=0.35$ from
Equation (6), we derive $R_{\rm gas}=R_{\rm gas*}\times
[M_h/(10^{12.3}\,\hmsol)]^{0.3}$ $h^{-1}$ kpc at $z\sim 0.4$.

The calculations above show that observations support the initial
assumptions in TC08 that the gaseous radius is a constant fraction of
the halo radius for all halos and that the gaseous radius scales with
halo mass according to $M_h^{t}$ with $t\approx 1/3$.  We note that
our galaxy sample therefore probes halo mass range from
$10^{10.5}\,\hmsol$ to $10^{12.8}\,\hmsol$, and that the best-fit
scaling relation is driven by galaxies in the cold-hot transition
regime.  Next, we compare the observed covering fraction of Mg\,II
absorbers with the predicted mean of TC08.

As summarized in \S\ 1, we find in the halo occupation analysis of
TC08 that dark matter halos of $M_h=10^{11.5-12.5}$ \hmsol\ on average
have a covering fraction of unity $\langle\kappa(M_h)\rangle=1$ at
$R_{\rm gas}\le R_{\rm 200}/3$ for Mg$^+$ and
$\langle\kappa(M_h)\rangle=0$ at larger radii.  This halo mass range
corresponds to a luminosity range of $L_B=0.2-1.4\,L_{B_*}$ at $z\sim
0.4$, following the mass-to-light scaling relation discussed above.
Sixteen galaxies in our sample have $L_B\ge 0.2\,L_{B_*}$, eleven of
which are at $\log\,\rho +0.14\times (M_B-M_{B_*}) < \log\,R_{\rm
gas*}$.  We identify a corresponding Mg\,II absorber for every one of
these galaxies, indicating a mean covering fraction of
$\langle\kappa\rangle=100$ \% at $R<R_{\rm gas}$.  This empirical
result agrees with our model expectation.  To assess the significance
of detecting a Mg\,II absorber in every galaxy in the sample of
eleven, we adopt the binomial distribution function and find that we
can constrain the underlying mean covering factor at
$\langle\kappa\rangle>0.76$ with greater than 95\% confidence level.

Three of the seven remaining $L_B<0.2\,L_{B_*}$ galaxies in our sample
have $\log\,\rho +0.14\times (M_B-M_{B_*}) < \log\,R_{\rm gas*}$.  Two
of them do not have corresponding Mg\,II absorbers identified to a
sensitive upper limit, implying $\langle\kappa\rangle=33$ \% at
$R<R_{\rm gas}$ for fainter galaxies (presumably lower-mass halos).
This is also consistent with the expectation of TC08, where we found a
quickly declining $\langle\kappa\rangle$ for halos of $M_h<10^{11.5}$
\hmsol, although the observations have a much lower statistical
significance.  In addition, we note that there exists an intrinsic
scatter in the fiducial scaling relation between $M_h$ and $L_B$.
This intrinsic scatter is expected to smear out the exact edge of
gaseous halos, particularly when considering an ensemble of objects
with a wide range of luminosity.

In summary, the initial results from the on-going survey of Mg\,II
absorbers in the vicinity of known galaxies close to QSO lines of
sight have allowed us to study in detail the properties of extended
gas around galaxies.  The best-fit scaling relation between the extent
of Mg$^+$ and galaxy luminosity and the observed mean gas covering
fraction together provide strong empirical evidence to support the
halo occupation model of TC08.  To improve the statistical
significances of various empirical parameters and to obtain a better
understanding of the incidence of cold gas in dark matter halos, such
as $\kappa$ vs.\ $M_h$ and $\kappa$ vs.\ $\rho$, a larger sample of
galaxy and Mg\,II absorber pairs established for a wide range of
galaxy luminosity and morphology is necessary.  In particular, only
three galaxies in our current sample would be considered in a
``cold-mode'' halo based on their luminosity and none shows a
corresponding Mg\,II absorber.  Deep surveys targeting at fainter
dwarfs are necessary to probe these ``cold-mode'' halos.  Finally, we
encourage observations of Mg\,II absorbers along the lines of sight
toward close QSO pairs for obtaining a direct measure of the incidence
of Mg$^+$ in individual halos.

\acknowledgments

  We thank J.-R.\ Gauthier, J.\ O'Meara, M.\ Rauch and G.\ Becker for
assistance on obtaining part of the MIKE spectra presented in this
paper.  We thank N.\ Gnedin and M.\ Rauch for helpful comments on an
early version of the paper.  This research has made use of the
NASA/IPAC Extragalactic Database (NED) which is operated by the Jet
Propulsion Laboratory, California Institute of Technology, under
contract with the National Aeronautics and Space Administration.
H.-W.C. acknowledges partial support from NASA Long Term Space
Astrophysics grant NNG06GC36G and an NSF grant AST-0607510.


\end{document}